\documentclass[prd,onecolumn,nofootinbib]{revtex4}
\usepackage{bm,amsmath,amssymb}

\begin{document}

\title{Torsional regularization of vertex function}
\author{Nikodem Pop{\l}awski}
\affiliation{Department of Mathematics and Physics, University of New Haven, 300 Boston Post Road, West Haven, CT 06516, USA}
\email{NPoplawski@newhaven.edu}

\begin{abstract}
The noncommutativity of the momentum components, arising from spacetime torsion coupled to spin, replaces the integration over the momentum in loop Feynman diagrams with the summation over the momentum eigenvalues.
This prescription regularizes logarithmically divergent integrals by turning them into convergent sums.
We apply torsional regularization to the vertex function in quantum electrodynamics in the one-loop correction.
The magnetic form factor is consistent with Schwinger's value for the anomalous magnetic dipole moment of a fermion.
We show that it depends on the mass of a fermion, which may explain the observed deviation of the magnetic moment of a muon from the predicted theoretical value.
We also show that the electric form factor is ultraviolet convergent.
\end{abstract}
\maketitle

\section{Torsional regularization}

Feynman diagrams in quantum field theory involve divergent integrals in the four-momentum space, arising from contributions with unbounded energy and momentum \cite{qft,PS}.
This situation is referred to as an ultraviolet divergence.
Since an infinite result is unphysical, it requires regularization: a mathematical method of modifying singular quantities and making them finite and physical.
The most common methods to eliminate an ultraviolet divergence are Pauli-Villars regularization \cite{PaVi} and dimensional regularization \cite{HoVe}.

In \cite{toreg}, we proposed a physical, torsional regularization, based on spacetime torsion \cite{Schr}.
The consistency of the conservation law for the total angular momentum of a Dirac particle in curved spacetime with relativistic quantum mechanics requires torsion \cite{req}.
We showed in \cite{toreg} that in the presence of the torsion tensor, the four-momentum components do not commute.
The quantum commutation relation for the four-momentum requires that the integration in the momentum space in Feynman diagrams must be replaced with the summation over the discrete momentum eigenvalues.
We used the simplest and most natural theory of gravity with torsion: the Einstein-Cartan theory, in which torsion is coupled to spin \cite{EC}.
We found a prescription for this summation that agrees with convergent integrals and showed that it regularizes logarithmically divergent integrals in loop diagrams, equivalently to Pauli-Villars regularization.
Torsional regularization, originating from quantization of momentum and spin-torsion coupling, resolves ultraviolet divergence similarly to how Planck's quantization of energy of emitted photons resolves the ultraviolet catastrophe in the black-body radiation.
In addition, torsion resolves the singularity problem in general relativity \cite{avert}, provides fermions with effective spatial extension \cite{nons}, and can explain the observed dynamics of the early Universe \cite{ApJ}.

The components $p_i$ of the four-dimensional momentum operator in the presence of torsion satisfy the commutation relation
\[
[p_i,p_j]=2i\hbar S^k_{\phantom{k}ij}p_k,
\]
where $S^k_{\phantom{k}ij}$ is the torsion tensor \cite{toreg}.
For the Dirac fields, they can be written as
\begin{equation}
[p_x,p_y] = iQp_z,\,\,\,[p_y,p_z] = iQp_x,\,\,\,[p_z,p_x] = iQp_y,
\end{equation}
where $Q=-(1/3)\hbar\epsilon^{0jkl}S_{jkl}$.
According to the Cartan equations, $Q$ depends on the four-momentum and is proportional to $p^3$, where $p^2=(p^0)^2-{\bf p}^2$ (we use $c=1$):
\begin{equation}
Q=Up^3,
\end{equation}
where $U$ is a constant whose unit is the same as that of the inverse squared momentum and whose value is on the order of the squared inverse of the Planck mass $M_\textrm{P}$.
Defining a vector
\begin{equation}
{\bf n}=\frac{{\bf p}}{Q}
\end{equation}
gives the cyclic commutation relations for its three spatial components $n_i$:
\begin{equation}
[n_x,n_y]=in_z,\,\,\,[n_y,n_z]=in_x,\,\,\,[n_z,n_x]=in_y.
\label{commut}
\end{equation}
Since these equations are analogous to the commutation relations for the angular momentum, the eigenvalues of ${\bf n}$ are given by $n=|{\bf n}|=\sqrt{j(j+1)}$ and $n_z=m$, where $j$ is the orbital quantum number (a nonnegative integer) and $m$ is the magnetic quantum number (an integer $m\in[-j,j]$) \cite{qm}.

In \cite{toreg}, we proposed that the integration over the ${\bf n}$ space satisfying (\ref{commut}) must be replaced with the summation over eigenstates:
\[
\int dn_x \int dn_y \int dn_z\, f(n)\rightarrow 4\pi \sum_{\textrm{eigenstates}} f(n)\,|n_z|,
\]
where $f(n)$ is any scalar function of $n$.
In this prescription, the integration over continuous space of $n_x$ and $n_y$ is replaced with the summation over the eigenvalues, the quantity under the sum is multiplied by the absolute value of the commutator of the integration variables, $|[n_x,n_y]|=|n_z|$, and the integration over $n_z$ is also replaced with the summation over the eigenvalues.
The summation gives \cite{toreg}
\begin{equation}
\int dn_x \int dn_y \int dn_z\, f(n)\rightarrow 4\pi\sum_{j=1}^\infty\sum_{m=-j}^j f(n)\,|m|=4\pi \sum_{j=1}^\infty f(n)\,j(j+1)=4\pi \sum_{j=1}^\infty f(n)\,n^2,
\label{pres}
\end{equation}
since for $j=0$ we have $m=0$ and thus $n_z=0$, which does not contribute to (\ref{pres}).

A typical loop Feynman diagram in quantum electrodynamics involves calculating an integral in the four-momentum space that has a form $\int d^4l/(l^2-\Delta+i\epsilon)^s$, where $\Delta>0$ does not depend on the four-momentum $l^i$, $s$ is an integer, and $\epsilon\to 0^{+}$ \cite{qft,PS}.
The integration is taken in spacetime with the Lorentzian metric signature $(+,-,-,-)$.
Calculations are simplified if we apply the Wick rotation in which the time component of the momentum $l^0$ is replaced with $il^0_\textrm{E}$.
Accordingly, $l^2_\textrm{E}=-[(l^0_\textrm{E})^2+{\bf l}^2_\textrm{E}]$ and the integration is taken in the four-dimensional Euclidean space with the metric signature $(+,+,+,+)$.
The integral becomes $i(-1)^s\int d^4l_\textrm{E}/(l^2_\textrm{E}+\Delta)^s=i(-1)^s\int dl_\textrm{E}^0 d{\bf l}/(l^2_\textrm{E}+\Delta)^s$, where $d{\bf l}=dl_x dl_y dl_z$.
Hereinafter, we omit the subscript E.
The relation $Q=Ul^3$ (without loss of generality, we assume $U>0$) gives
\begin{equation}
l^2=(l^0)^2+U^2 n^2 l^6.
\label{condit}
\end{equation}
The integration over $dl^0$ can thus be replaced with the integration over $dl$:
\[
dl^0=dl \frac{dl^0}{dl}=dl \frac{1-3U^2 n^2 l^4}{(1-U^2 n^2 l^4)^{1/2}}.
\]

The integration over the ${\bf l}$ space can be replaced with the integration over the ${\bf n}$ space and then with the summation over $j$:
\begin{equation}
\int dl_x \int dl_y \int dl_z\, f({\bf l}^2)\rightarrow \int dn_x \int dn_y \int dn_z J\,f(Q^2 n^2)\rightarrow 4\pi \sum_{j=1}^\infty J\,f(Q^2 n^2)\,n^2,
\label{prescr}
\end{equation}
where $J=\partial(l_x,l_y,l_z)/\partial(n_x,n_y,n_z)$ is the Jacobian of the transformation from the components of ${\bf l}$ to the components of ${\bf n}$.
Differentiating (\ref{condit}) with respect to $n_x$, using $n^2=n_x^2+n_y^2+n_z^2$, gives $2l\frac{\partial l}{\partial n_x}=6U^2 n^2 l^5 \frac{\partial l}{\partial n_x}+2U^2 l^6 n_x$ and thus
\[
\frac{\partial l}{\partial n_x}=\frac{U^2 l^5 n_x}{1-3U^2 n^2 l^4}.
\]
Consequently, we find
\begin{eqnarray}
& & \frac{\partial l_x}{\partial n_x}=\frac{\partial(Qn_x)}{\partial n_x}=Q+3Un_x l^2 \frac{\partial l}{\partial n_x}=\frac{Q}{1-3U^2 n^2 l^4}[1-3U^2 l^4 (n_y^2 + n_z^2)], \nonumber \\
& & \frac{\partial l_x}{\partial n_y}=\frac{\partial(Qn_x)}{\partial n_y}=3Un_x l^2 \frac{\partial l}{\partial n_y}=\frac{Q}{1-3U^2 n^2 l^4}(3U^2 l^4 n_x n_y), \nonumber
\end{eqnarray}
and similarly for other components.
The Jacobian is
\[
J=\mbox{det}\left( \begin{array}{ccc}
\partial l_x/\partial n_x & \partial l_x/\partial n_y & \partial l_x/\partial n_z \\
\partial l_y/\partial n_x & \partial l_y/\partial n_y & \partial l_y/\partial n_z \\
\partial l_z/\partial n_x & \partial l_z/\partial n_y & \partial l_z/\partial n_z \end{array} \right)=\frac{Q^3}{1-3U^2 n^2 l^4}.
\]
Therefore, the integration over the Euclidean four-momentum $l^i$ and using the prescription ({\ref{prescr}) gives
\begin{eqnarray}
& & \int dl^0 d{\bf l}=\int dl \frac{dl^0}{dl} J\,d{\bf n}=\int dl\,d{\bf n}\frac{Q^3}{(1-U^2 n^2 l^4)^{1/2}}=2\int_0^{\sqrt{1/(Un)}}dl\,d{\bf n}\frac{U^3 l^9}{(1-U^2 n^2 l^4)^{1/2}} \nonumber \\
& & \rightarrow 8\pi \sum_{j=1}^\infty\int_0^{\sqrt{1/(Un)}}dl \frac{U^3 l^9}{(1-U^2 n^2 l^4)^{1/2}} n^2, \nonumber
\end{eqnarray}
with $n=\sqrt{j(j+1)}$.

Finally, the integral $\int dl^0 d{\bf l}/(l^2+\Delta)^s$ in the noncommutative momentum space resulting from torsion coupled to spin is \cite{toreg}
\begin{eqnarray}
& & \int dl^0 d{\bf l}\frac{1}{(l^2+\Delta)^s}\rightarrow 8\pi \sum_{j=1}^\infty\int_0^{\sqrt{1/(Un)}}dl \frac{U^3 l^9}{(1-U^2 n^2 l^4)^{1/2}} n^2\frac{1}{(l^2+\Delta)^s}=8\pi \sum_{j=1}^\infty\int_0^1 d\xi \frac{U^3 \xi^9 n^2 (Un)^{-5}}{(1-\xi^4)^{1/2}[\xi^2/(Un)+\Delta]^s} \nonumber \\
& & =\sum_{j=1}^\infty\int_0^1 d\xi \frac{8\pi U^{-2} \xi^9 n^{-3} (Un)^s}{(1-\xi^4)^{1/2}(\xi^2+U\Delta n)^s}=\sum_{j=1}^\infty\int_0^1 d\zeta \frac{4\pi U^{s-2} \zeta^4 n^{s-3}}{(1-\zeta^2)^{1/2}(\zeta+U\Delta n)^s}=4\pi \sum_{j=1}^\infty\int_0^{\pi/2} d\phi \frac{U^{s-2} \sin^4\phi\,n^{s-3}}{(\sin\phi+U\Delta n)^s} \nonumber \\
& & =4\pi \sum_{j=1}^\infty\int_0^{\pi/2} d\phi \frac{U^{s-2} \sin^4\phi\,(j(j+1))^{(s-3)/2}}{[\sin\phi+U\Delta (j(j+1))^{1/2}]^s},
\label{sumint}
\end{eqnarray}
where we denote $Unl^2=\xi^2=\zeta=\sin\phi$.
Since $U>0$, at large values of $j$ this integral behaves as $\sim\sum_{j=1}^\infty j^{-3}$ for any value of $s$.
The sum-integral (\ref{sumint}) thus converges and torsional regularization works for any $s$.

In the limit of continuous momentum space, $U\rightarrow 0$, the separation between adjacent values of $j$ does not affect significantly the integrand and thus the summation over $j$ can be replaced with the integration over $y$, where $y=\sqrt{j(j+1)}$.
In \cite{toreg} we showed that for $s=3$, this limit gives an ordinary, finite value of the integral.
For $s=2$, the integral in this limit diverges as $\sim\ln(U)$, which is equivalent to Pauli-Villars regularization.

\section{Magnetic form factor and anomalous magnetic moment}

In quantum electrodynamics, the coupling between an electron and a photon in the leading order of perturbation theory is given by a term proportional to the Dirac matrix $\gamma^\mu$: $-ie\gamma^\mu$, where $e$ is the electric charge of an electron and we use $\hbar=1$ \cite{qft,PS}.
Beyond the leading order, this coupling is given by the vertex function $\Gamma^\mu=\gamma^\mu+\delta\Gamma^\mu$.
The one-loop correction to the vertex function $\delta\Gamma^\mu$ describes a process in which an electron emits a (virtual) photon, emits a second photon, and then reabsorbs the first.
This correction is the dominant contribution to the anomalous magnetic dipole moment of an electron.

The one-loop correction to the vertex function is given by \cite{PS}
\begin{eqnarray}
& & \delta\Gamma^\mu=2ie^2\int\frac{d^4 l}{(2\pi)^4}\int_0^1 dx\,dy\,dz\,\delta(x+y+z-1)\frac{2}{(l^2-\Delta+i\epsilon)^3} \nonumber \\
& & \cdot\Bigl[\gamma^\mu\Bigl(-\frac{1}{2}l^2+(1-x)(1-y)q^2+(1-4z+z^2)m^2\Bigr)+\frac{i\sigma^{\mu\nu}q_\nu}{2m}\bigl(2m^2 z(1-z)\bigr)\Bigr], \nonumber
\end{eqnarray}
where $l^2=l_\mu l^\mu$, $l^\mu=(k+yq-zp)^\mu$, $p^\mu$ is the four-momentum of the incoming electron, $q^\mu$ is the incoming four-momentum of the external photon, $(p-k)^\mu$ is the four-momentum of the virtual photon, $\sigma^{\mu\nu}=(i/2)[\gamma^\mu,\gamma^\nu]$, $m$ is the mass of an electron, $x,y,z$ are the Feynman parameters (in the range $[0,1]$ and satisfying $x+y+z=1$), and
\[
\Delta=(1-z)^2 m^2-xyq^2\ge 0.
\]
This function can be written as $\Gamma^\mu=\gamma^\mu F_1(q^2)+i\sigma^{\mu\nu}q_\nu/(2m) F_2(q^2)$, where $F_1(q^2)$ and $F_2(q^2)$ are form factors that depend only on the square of the four-momentum of the photon $q^2$.
Accordingly, the form factors are \cite{PS}
\begin{eqnarray}
& & F_1(q^2)=1-2ie^2\int_0^1 dx\,dy\,dz\,\delta(x+y+z-1)\int\frac{d^4 l}{(2\pi)^4}\frac{N}{(l^2-\Delta+i\epsilon)^3}, \label{form1} \\
& & F_2(q^2)=2ie^2\int_0^1 dx\,dy\,dz\,\delta(x+y+z-1)\int\frac{d^4 l}{(2\pi)^4}\frac{4m^2 z(1-z)}{(l^2-\Delta+i\epsilon)^3}, \label{form2}
\end{eqnarray}
where
\[
N=l^2-2(1-x)(1-y)q^2-2(1-4z+z^2)m^2.
\]

We consider the magnetic form factor (\ref{form2}) at $q^2=0$, which corresponds to the anomalous magnetic dipole moment of an electron: $(g-2)/2=F_2(0)$, where $g$ is the gyromagnetic ratio.
For $q^2=0$, $\Delta=(1-z)^2 m^2$ and the integration over $x,y$ is trivial:
\begin{eqnarray}
& & F_2(0)=2ie^2\int_0^1 dz(1-z)\int\frac{d^4 l}{(2\pi)^4}\frac{4m^2 z(1-z)}{(l^2-\Delta+i\epsilon)^3}=\frac{2i\alpha m^2}{\pi^3}\int_0^1 dz\int d^4 l\frac{z(1-z)^2}{(l^2-\Delta+i\epsilon)^3} \nonumber \\
& & =\frac{2\alpha m^2}{\pi^3}\int_0^1 dz\int d^4 l_\textrm{E}\frac{z(1-z)^2}{(l^2_\textrm{E}+\Delta)^3}, \nonumber
\end{eqnarray}
where $\alpha=e^2/(4\pi)$ is the fine-structure constant.
Omitting the subscript E, this integral is
\begin{equation}
F_2(0)=\frac{2\alpha m^2}{\pi^3}\int_0^1 dz\,z(1-z)^2\int dl^0 d{\bf l}\frac{1}{(l^2+\Delta)^3},
\label{magnet}
\end{equation}
which has the form (\ref{sumint}) for $s=3$.
Analogously to (\ref{sumint}), torsional regularization of this integral gives
\begin{eqnarray}
& & F_2(0)\rightarrow \frac{2\alpha m^2}{\pi^3}\int_0^1 dz\,z(1-z)^2 4\pi \sum_{j=1}^\infty\int_0^{\pi/2} d\phi \frac{U^{s-2} \sin^4\phi\,(j(j+1))^{(s-3)/2}}{[\sin\phi+U\Delta (j(j+1))^{1/2}]^s} \nonumber \\
& & =\frac{8\alpha m^2 U}{\pi^2}\sum_{j=1}^\infty \int_0^1 dz\,z(1-z)^2 \int_0^{\pi/2} d\phi \frac{\sin^4\phi}{[\sin\phi+m^2 U\,(j(j+1))^{1/2} (1-z)^2]^3}.
\label{magnetic}
\end{eqnarray}

The sum-integral (\ref{magnetic}) depends only on the nondimensional quantity
\[
\chi=m^2 U.
\]
Replacing $z$ with $1-z$, we obtain
\[
F_2(0)=\frac{8\alpha\chi}{\pi^2}\sum_{j=1}^\infty \int_0^{\pi/2} d\phi\,\sin^4\phi \int_0^1 dz\frac{(1-z)z^2}{[\sin\phi+\chi\,(j(j+1))^{1/2} z^2]^3}.
\]
Defining
\[
b=\frac{\sin\phi}{\chi\sqrt{j(j+1)}}
\]
and $u=z/\sqrt{b}$, we obtain
\begin{eqnarray}
& & F_2(0)=\frac{8\alpha\chi}{\pi^2}\sum_{j=1}^\infty \int_0^{\pi/2} d\phi\,\sin\phi \int_0^1 dz\frac{(1-z)z^2}{(1+z^2/b)^3} \nonumber \\
& & =\frac{8\alpha\chi}{\pi^2}\sum_{j=1}^\infty \int_0^{\pi/2} d\phi\,\sin\phi\Bigl[\frac{b^{3/2}}{8}\Bigl(\frac{u(u^2-1)}{(u^2+1)^2}+\mbox{arc tan}\,u\Bigr)+\frac{b^2(2u^2+1)}{4(u^2+1)^2}\Bigr]\Big|_{u=0}^{u=1/\sqrt{b}} \nonumber \\
& & =\frac{\alpha\chi}{\pi^2}\sum_{j=1}^\infty \int_0^{\pi/2} d\phi\,\sin\phi\Bigl[b^{3/2}\mbox{arc tan}(b^{-1/2})-\frac{b^2}{1+b}\Bigr].
\label{anomal}
\end{eqnarray}
This sum-integral (\ref{anomal}) is finite because the sum over $j$ converges as $\sim\sum_{j=1}^\infty j^{-3/2}$.
It gives the anomalous magnetic dipole moment of a charged fermion with mass $m$ at the one-loop order in quantum electrodynamics.

Since $U\sim M^{-2}_\textrm{P}$, $\chi$ is on the order of $(m/M_\textrm{P})^2\ll 1$.
Therefore, we can consider the limit of continuous momentum space, $U\to 0$ ($\chi\to 0$).
In this limit, the separation between adjacent values of $j$ does not affect significantly the integrand and thus the summation over $j$ can be replaced with the integration over $n$, where $n=\sqrt{j(j+1)}$.
We obtain
\begin{eqnarray}
& & \lim_{\chi\to 0} F_2(0)=\lim_{\chi\to 0} \frac{8\alpha\chi}{\pi^2}\sum_{j=1}^\infty \int_0^1 dz\,z(1-z)^2 \int_0^{\pi/2} d\phi \frac{\sin^4\phi}{[\sin\phi+\chi(1-z)^2 (j(j+1))^{1/2}]^3} \nonumber \\
& & =\lim_{\chi\to 0} \frac{8\alpha\chi}{\pi^2}\int_C^\infty dn \int_0^{\pi/2} d\phi\,\sin^4\phi \int_0^1 dz\frac{(1-z)z^2}{[\sin\phi+\chi z^2 n]^3} \nonumber \\
& & =\lim_{\chi\to 0} \frac{8\alpha\chi}{\pi^2}\int_0^{\pi/2} d\phi\,\sin^4\phi \int_0^1 dz(1-z)z^2 \int_0^\infty\frac{dw}{\chi z^2 (\sin\phi+w)^3} \nonumber \\
& & =\frac{8\alpha}{\pi^2}\int_0^{\pi/2} d\phi\,\sin^4\phi \int_0^1 dz(1-z)\frac{1}{2\sin^2\phi}=\frac{\alpha}{2\pi},
\label{qed}
\end{eqnarray}
where $w=\chi z^2 n$.
The exact value of the lower limit $C>0$ is irrelevant because the change of the integration variable turns the lower limit into zero.

The result (\ref{qed}) is the limit of continuous momentum space for the anomalous magnetic dipole moment (\ref{magnetic}) of a charged fermion at the one-loop order in quantum electrodynamics and agrees with the famous result found by Schwinger \cite{PS,Schw}.
Since $\chi\ll 1$, the equivalent values (\ref{magnetic}) and (\ref{anomal}) do not significantly differ from $\alpha/(2\pi)$, which can be verified numerically.
In order to obtain a more accurate value for the anomalous magnetic moment of a fermion, one must consider higher-order corrections to the vertex function and apply torsional regularization at each order, replacing the integration over the momentum with the summation over its eigenstates.

The sum-integral (\ref{magnetic}) depends on the mass of an electron.
Consequently, the anomalous magnetic moment of a muon at the one-loop order in quantum electrodynamics is different from that of an electron.
This difference arises from the noncommutativity of the momentum, resulting from torsion coupled to spin.
This difference, extended to the weak interaction, may explain the observed deviation of the measurement of the magnetic moment of a muon from the theoretical value predicted by the Standard Model of particle physics \cite{obs}.

\section{Regularization of electric form factor}

We consider the electric form factor (\ref{form1}) at $q^2=0$, which describes the one-loop correction to the electric charge of an electron in the large-distance limit.
For $q^2=0$, we obtain
\begin{eqnarray}
& & F_1(0)-1=-2ie^2\int_0^1 dz(1-z)\int\frac{d^4 l}{(2\pi)^4}\frac{l^2-2(1-4z+z^2)m^2}{(l^2-\Delta+i\epsilon)^3}=-\frac{i\alpha}{2\pi^3}\int_0^1 dz(1-z)\int d^4 l\frac{l^2-2(1-4z+z^2)m^2}{(l^2-\Delta+i\epsilon)^3} \nonumber \\
& & =\frac{\alpha}{2\pi^3}\int_0^1 dz(1-z)\int d^4 l_\textrm{E}\frac{l^2_\textrm{E}+2(1-4z+z^2)m^2}{(l^2_\textrm{E}+\Delta)^3}. \nonumber
\end{eqnarray}
Omitting the subscript E, this integral is
\begin{eqnarray}
& & F_1(0)-1=\frac{\alpha}{2\pi^3}\int_0^1 dz(1-z)\int dl^0 d{\bf l}\frac{l^2+\Delta+(1-6z+z^2)m^2}{(l^2+\Delta)^3} \nonumber \\
& & =\frac{\alpha}{2\pi^3}\int_0^1 dz(1-z)\int dl^0 d{\bf l}\frac{1}{(l^2+\Delta)^2}+\frac{\alpha m^2}{2\pi^3}\int_0^1 dz(1-z)(1-6z+z^2)\int dl^0 d{\bf l}\frac{1}{(l^2+\Delta)^3}.
\label{electric}
\end{eqnarray}

The first term in (\ref{electric}) has the form (\ref{sumint}) for $s=2$.
Analogously to (\ref{sumint}), torsional regularization of this integral gives
\begin{eqnarray}
& & F_1(0)_\textrm{first}\rightarrow \frac{\alpha}{2\pi^3}\int_0^1 dz(1-z)\,4\pi \sum_{j=1}^\infty\int_0^{\pi/2} d\phi \frac{U^{s-2} \sin^4\phi\,(j(j+1))^{(s-3)/2}}{[\sin\phi+U\Delta (j(j+1))^{1/2}]^s} \nonumber \\
& & =\frac{2\alpha}{\pi^2}\sum_{j=1}^\infty \int_0^1 dz(1-z) \int_0^{\pi/2} d\phi \frac{\sin^4\phi\,(j(j+1))^{-1/2}}{[\sin\phi+m^2 U\,(j(j+1))^{1/2} (1-z)^2]^2}. \nonumber
\end{eqnarray}
Following the steps leading to (\ref{anomal}), we obtain
\begin{eqnarray}
& & F_1(0)_\textrm{first}=\frac{2\alpha}{\pi^2}\sum_{j=1}^\infty \int_0^{\pi/2} d\phi\,\sin^4\phi \int_0^1 dz\frac{z(j(j+1))^{-1/2}}{[\sin\phi+\chi\,(j(j+1))^{1/2} z^2]^2} \nonumber \\
& & =\frac{2\alpha}{\pi^2}\sum_{j=1}^\infty \int_0^{\pi/2} d\phi\,\sin^2\phi \int_0^1 dz\frac{z(j(j+1))^{-1/2}}{(1+z^2/b)^2}=\frac{\alpha}{\pi^2}\sum_{j=1}^\infty \int_0^{\pi/2} d\phi\,\sin^2\phi \frac{b}{1+b} (j(j+1))^{-1/2} \nonumber \\
& & =\frac{\alpha}{\pi^2}\sum_{j=1}^\infty \int_0^{\pi/2} d\phi\,\sin^3\phi \frac{1}{\sin\phi(j(j+1))^{1/2}+\chi\,j(j+1)}.
\label{first}
\end{eqnarray}
Since $\chi>0$, the sum-integral (\ref{first}) converges as $\sim\sum_{j=1}^\infty j^{-2}$.
Without torsion ($\chi=0$), this sum-integral would diverge as $\sim\sum_{j=1}^\infty j^{-1}$.
As $\chi\to 0$, that divergence is $\sim\ln(\chi)$.

The second term in (\ref{electric}) has the form (\ref{sumint}) for $s=3$.
However, this term has an infrared divergence arising from $\Delta=0$ if $z=1$ (it occurs for $s=3$ but not for $s=2$).
This divergence, related to a virtual soft photon in the vertex function, is canceled by the contributions from braking radiation (bremsstrahlung) involving soft photons.
Mathematically, this cancelation can be carefully treated by adding to $\Delta$ a term $z\mu^2$ which represents a fictitious, small, nonzero mass $\mu$ of the virtual photon \cite{PS}, and then taking the limit $\mu\to 0$.

Analogously to (\ref{magnet}) and (\ref{magnetic}), torsional regularization of this integral gives
\begin{eqnarray}
& & F_1(0)_\textrm{second}\rightarrow \frac{\alpha m^2}{2\pi^3}\int_0^1 dz(1-z)(1-6z+z^2) 4\pi \sum_{j=1}^\infty\int_0^{\pi/2} d\phi \frac{U^{s-2} \sin^4\phi\,(j(j+1))^{(s-3)/2}}{[\sin\phi+U(\Delta+z\mu^2) (j(j+1))^{1/2}]^s} \nonumber \\
& & =\frac{2\alpha m^2 U}{\pi^2}\sum_{j=1}^\infty \int_0^1 dz(1-z)(1-6z+z^2) \int_0^{\pi/2} d\phi \frac{\sin^4\phi}{[\sin\phi+(m^2 U(1-z)^2 +\mu^2 Uz)(j(j+1))^{1/2}]^3}. \nonumber
\end{eqnarray}
Following the steps leading to (\ref{anomal}), we obtain
\begin{eqnarray}
& & F_1(0)_\textrm{second}=\frac{2\alpha\chi}{\pi^2}\sum_{j=1}^\infty \int_0^{\pi/2} d\phi\,\sin^4\phi \int_0^1 dz\frac{z(z^2+4z-4)}{[\sin\phi+\chi\,(j(j+1))^{1/2} z^2 +\mu^2 U\,(j(j+1))^{1/2} (1-z)]^3} \nonumber \\
& & =\frac{2\alpha\chi}{\pi^2}\sum_{j=1}^\infty \int_0^{\pi/2} d\phi\,\sin\phi \int_0^1 dz\frac{z(z^2+4z-4)}{(1+z^2/b+(1-z)/a)^3},
\label{second}
\end{eqnarray}
where
\[
a=\frac{\sin\phi}{\mu^2 U\sqrt{j(j+1)}}.
\]
It can be shown that the sum-integral (\ref{second}) is also finite.
The electric form factor at $q^2=0$ is
\[
F_1(0)=1+F_1(0)_\textrm{first}+F_1(0)_\textrm{second}.
\]
A more extensive analysis of eliminating infrared divergence in Feynman diagrams that are already ultraviolet convergent (due to torsional regularization) will be considered elsewhere.

This work was funded by the University Research Scholar program at the University of New Haven.

\end{document}